\documentclass[preprint,showpacs,amsmath,amssymb]{revtex4}

  \usepackage[dvips]{graphicx}

\usepackage{dcolumn}
\usepackage{bm}

\begin{document}

\title{Field-controlled suppression of phonon-induced transitions
in coupled quantum dots}

\author{Andrea Bertoni}
\affiliation{INFM National Research Center on nanoStructures
and bioSystems at Surfaces ($S3$), Via Campi 213/A, 41100
Modena, Italy}
\author{Massimo Rontani}
\affiliation{INFM National Research Center on nanoStructures
and bioSystems at Surfaces ($S3$), Via Campi 213/A, 41100
Modena, Italy}
\author{Guido Goldoni}
\affiliation{INFM National Research Center on nanoStructures
and bioSystems at Surfaces ($S3$), 
Modena, Italy}
\affiliation{Dipartimento di
Fisica, Universit\`a di Modena e Reggio Emilia, Italy}
\author{Filippo Troiani}
\affiliation{INFM National Research Center on nanoStructures
and bioSystems at Surfaces ($S3$), 
Modena, Italy}
\affiliation{Dipartimento di
Fisica, Universit\`a di Modena e Reggio Emilia, Italy}
\author{Elisa Molinari}
\affiliation{INFM National Research Center on nanoStructures
and bioSystems at Surfaces ($S3$), 
Modena, Italy}
\affiliation{Dipartimento di
Fisica, Universit\`a di Modena e Reggio Emilia, Italy}

\date{\today}

\begin{abstract}
We   suggest    that      order-of-magnitude reduction      of     the
longitudinal-acoustic phonon scattering rate, the dominant decoherence
mechanism in quantum  dots, can be achieved  in coupled structures  by
the application of an external electric or magnetic field.  Modulation
of the    scattering rate  is  traced  to    the relation  between the
wavelength  of the emitted phonon  and the length scale of delocalized
electron wavefunctions.  Explicit   calculations   for 
realistic devices,  performed with a Fermi  golden rule approach and a
fully three-dimensional description of  the electronic quantum states,
show   that the  lifetime of   specific states  can  achieve tens of
$\mu$s.  Our findings extend the  feasibility basis of many  proposals
for quantum gates based on coupled quantum dots.
\end{abstract}


\maketitle

The loss of  quantum coherence, brought  about by the coupling between
charge  degrees  of freedom  and  lattice  vibrations, constitute a main
limitation for  single   and  coupled  quantum  dot (SQD   and   CQD)
implementations of  quantum  logic gates, as  well  as for a  wide class of
novel few- and single-electron devices. In quantum dots (QDs)
\cite{dotsrev}  the discrete nature
of the energy   spectrum  strongly    reduces this
interaction  with   respect  to   other semiconductor    systems.   In
particular, the coupling  with the dispersionless optical phonons, the
main source of decoherence  in structures of higher dimensionality, is
either  completely  suppressed   \cite{bockelmann90} or, in   specific
conditions,     leads     to      localized        polaronic    states
\cite{inoshita97hameau99}. Still,  however, emission  of longitudinal
acoustic (LA) phonons \cite{jacak03} limits the charge coherence times
to  tens    of ns in  typical    devices,  since intra-band electronic
transitions  merge  into the    LA  phonon continuum.    Tailoring the
coupling between  electronic quantum states and  LA phonons would be a
fundamental step toward the   design of quantum devices   with optimal
operation conditions.

In   this Letter we show that    LA-phonon scattering rate of specific
transitions can  be   suppressed in CQDs  by  proper  structure design
(as already discussed in Ref.~\onlinecite{zanardi98} for quantum dot arrays)
and by  use of  external  magnetic  and electric  fields.  This  is
accomplished by exploiting the interplay between the wavevector of the
emitted or absorbed phonon and the length scale  of the quantum states.
While the  above  effect  is    small in    typical SQDs,  which   are
characterized   by      a  single   length scale,     we    show   that
order-of-magnitude modulation of the scattering  rate can be achieved
in coherently coupled QDs,  where two characteristic length scales come
into play, namely the dimension  of each dot and  the dimension of the
CQD system as a whole.

Decay times of a single electron in  a SQD have been measured recently
by transport spectroscopy  \cite{fujisawa02}. These experiments, which
specifically probe the decay of carriers injected in the first excited
state to  the ground  state,  showed a  satisfactory agreement with  a
Fermi golden rule estimate of the electron relaxation times.  Below we
investigate  the decay time of   specific low energy transitions which
are, therefore, directly accessible with similar techniques.
As a  prototypical  system, we  consider a  GaAs/AlGaAs CQD  structure
formed by two identical QDs with cylindrical shape, coupled along the
growth direction \cite{tarucha96rontani04}.
We  describe the electronic quantum states
within  the familiar envelope   function approximation and  consider a
parabolic confinement in  the   $xy$ plane  of  the cylinders,  and  a
symmetric double-well in the $z$ direction: $V({\bf r})= V_z(z) +
\frac{1}{2}m\omega_0^2(x^2+y^2)$ where $V_z(z)=  V_l$ if  $L_b/2
\leq |z| \leq (L_b/2 + L_d)$ and $V_z(z)= V_h$ otherwise. Here
$L_d$ is the thickness of the  GaAs layers and  $L_b$ is the thickness
of the  inter-dot layer.  $(V_h-V_l)$ is the band-offset of GaAs/AlGaAs.
Furthermore, we consider either a magnetic
($B$) or  an electric ($E$)  field applied in  the $z$ direction.  For
this  cylindrically    symmetric   configuration,   the eigenfunctions
can be  given  the separable   form
$\psi_{nmg}({\bf        r})=  \phi_{nm}(x,y)   \chi_{g}(z)$,  with
$n=0,1,\dots$ the radial  and  $m=0,\pm 1,\dots$ the  angular  quantum
number of  the  Fock-Darwin   state  $\phi_{nm}(x,y)$;  $g=0$  ($1$)
indicates the ground   (first   excited) eigenstate for  the    biased
double-well potential along  the $z$ direction.

We use  the standard deformation-potential   model for the electron-LA
phonon  interaction,   with    the  interaction    Hamiltonian   ${\bf
H}_{e-p}=\sum_{\bf q} F(q)  (  b_{\bf q}  e^{i{\bf  q r}} + b_{\bf  q}
^{\dagger} e^{-i{\bf   q r}  })  $,  where   $b_{\bf q}$  and  $b_{\bf
q}^{\dagger}$ are the   annihilation   and creation operators   for  a
LA-phonon with wavevector ${\bf q}$ and $F(q)= q
\sqrt{{\hbar  D^2}/{(2\rho\omega_qV)}}$. $D$ is the deformation
potential, $\rho$ is the crystal  density  and $V$  the volume of  the
system.  A linear dispersion approximation $\omega_q= v q$ is used for
the LA branch,   with $v$ longitudinal sound   speed \cite{footnote1}.
Furthermore,  we assume that the device  operates in a single-electron
regime: electron-electron scattering is neglected accordingly.

Below we shall concentrate   on the scattering rate  between  specific
excited (initial) states $\psi_i$,  with energy $E_i$, and the  ground
(final) state $\psi_f$,    with energy $E_f$ \cite{footnote2}.
The scattering rate at zero temperature, given by the Fermi golden  rule, is
$\, \tau^{-1}=\frac{2\pi}{\hbar}\sum_{\bf q} |F(q)M_{fi}({\bf q})|^2
\delta(E_f-E_i+\hbar\omega_{\bf q}) \, , $
where
$M_{fi}({\bf  q})=\langle \psi_f |  e^{-i{\bf q r}  } | \psi_i\rangle$.

It is  convenient  to  write the  phonon  wave  vector  ${\bf  q}$  in
spherical coordinates and separate  the scattering matrix
$M_{fi}({\bf q})$ into a $z$ component $M_{fi}^{(z)}(q_z=q
\cos\theta)$,  to be evaluated numerically, and an in-plane
component $M_{fi}^{(xy)}(q_x=q \cos\varphi\sin\theta,q_y=q
\sin\varphi\sin\theta)$ which can  be  exactly     derived
\cite{bockelmann94}.   Performing    the integration over $q$, one
finally obtains:
\begin{eqnarray}\label{fermigoldenrule2}
\tau^{-1}=\frac{q_0^2} {v(2\pi\hbar)^2} F(q_0)
\int_0^{2\pi}\!d\varphi \int_0^{\pi/2}\! d\theta\sin\theta
                                                        \nonumber    \\
\times |M_{fi}^{(z)}(q_0 \cos\theta) M_{fi}^{(xy)}
(q_0 \cos\varphi\sin\theta,q_0
\sin\varphi\sin\theta)|^2 ,
\end{eqnarray}
with $q_0=(E_i-E_f)/(\hbar v)$. We emphasize that the decoupling of the
in-plane      and     vertical       degrees     of    freedom      in
Eq.~(\ref{fermigoldenrule2}) is approximately   valid as long  as  the
in-plane and vertical  confinement have different length scales,  as is
achieved by many growth techniques.

\begin{figure}
\includegraphics[width=.98\linewidth]{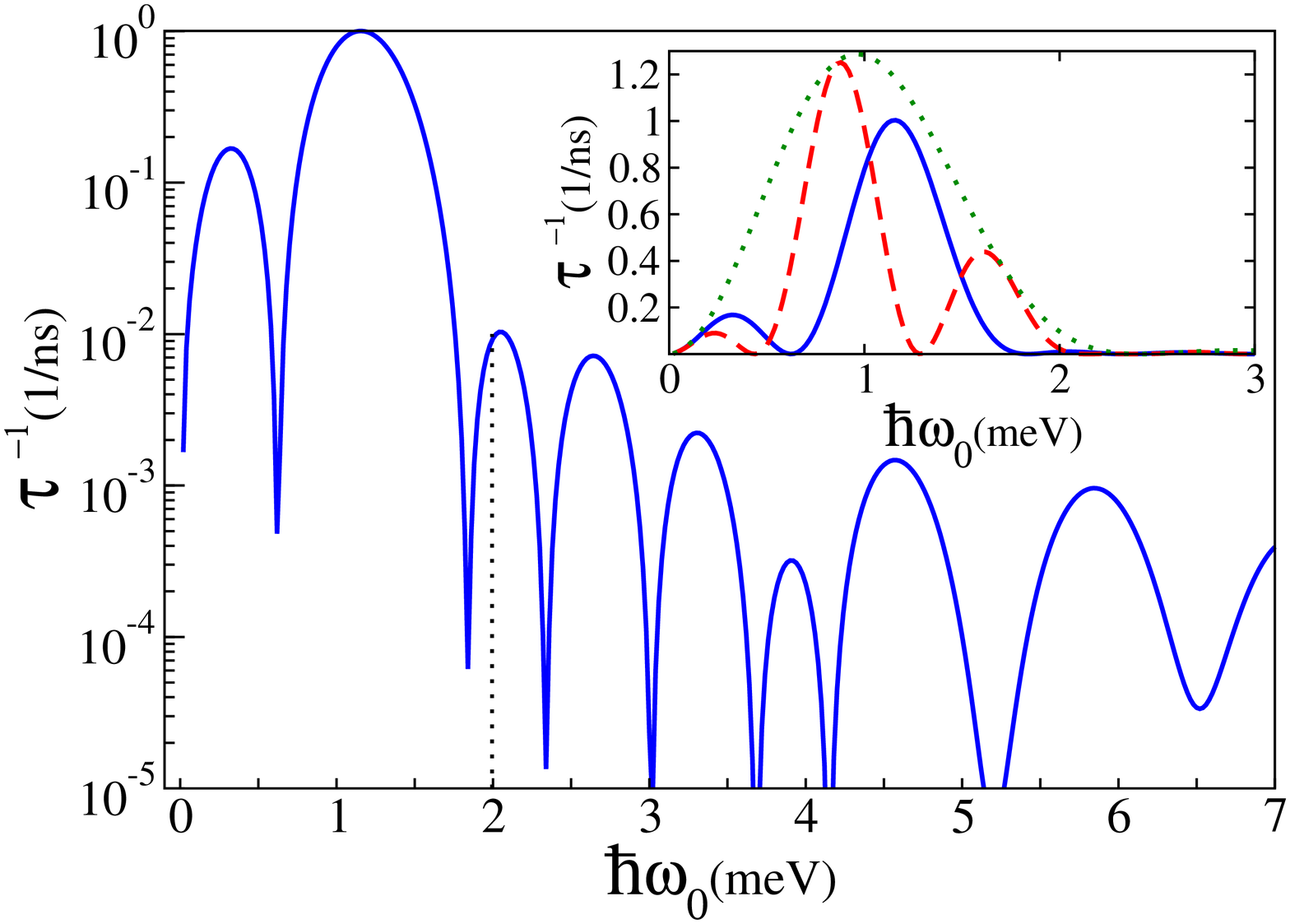}
\caption{\label{fig1}}
\end{figure}

Now    we  focus on   the    $(n,m,g) = (0,-1,0)  \rightarrow
(0,0,0)$ transition,  which  only involves in-plane    degrees of
freedom.   We choose  a CQD system  (see  Fig.~\ref{fig1},
caption) such that the tunneling energy is larger than
$\hbar\omega_0$ and $(0,-1,0)$ is the lowest excited state.
Figure~\ref{fig1} shows  the computed scattering rate  at 
$B=0$~T and zero temperature as  a function of the  in-plane confining energy
$\hbar \omega_0$.  The striking oscillations, which  cover several
orders  of magnitude, have their origin in the phase relation
between the phonon plane wave corresponding to the  considered
electronic transition and the electron wavefunction delocalized
between the two    QDs:  two limiting conditions, one with 
maximum overlap, the  other  with minimum overlap, due to electron-phonon
antiphase, occur  repeatedly  as the confinement energy (and,
therefore, the energy of the matching phonon) is varied; the two
cases  correspond  to a maximum and a minimum of the matrix
element $M_{fi}^{(z)}$, respectively.
In the inset of Fig.~\ref{fig1} we compare the  scattering rate of
two CQD structures,  with different barrier  widths, with the one
of a SQD.
As one may expect, the curves have  a similar limiting behavior:
when the confining potential and, as a  consequence, the energy of
the  emitted phonon is   very  low, $\tau^{-1}$ decreases due  to
(a) the $q^2$ prefactor ensuing from the low-energy  LA-phonon
density of states and the linear  $q$ dependence of $F(q)$  in
Eq.~(\ref{fermigoldenrule2}), and (b) the decreasing  value  of
$M^{(xy)}_{fi}({\bf  q})$ due  to orthogonality of  the  states
(at vanishing  $q$  the electron-phonon Hamiltonian   becomes
constant).  In    the opposite limit, for $\hbar\omega_0>$~2~meV,
the phonon oscillation
length becomes  much smaller than the typical length scale of  the
electron wave function;  as  a consequence  $M_{fi}({\bf q})$
vanishes since the initial and final electron states do not posses
the proper Fourier component able to trigger  the phonon
oscillation.  In the  {\it intermediate regime},  while the  SQD
curve  shows  a single well-defined maximum,  the  CQD  curve
oscillates.    The oscillation pattern depends on the geometry of
the system \cite{zanardi98}. If the distance between the dots is
increased, the energy difference  between the maxima     decreases
(see inset in  Fig.~\ref{fig1}),  with the CQD curve enveloped by
the SQD curve.

\begin{figure}
\includegraphics[width=.98\linewidth]{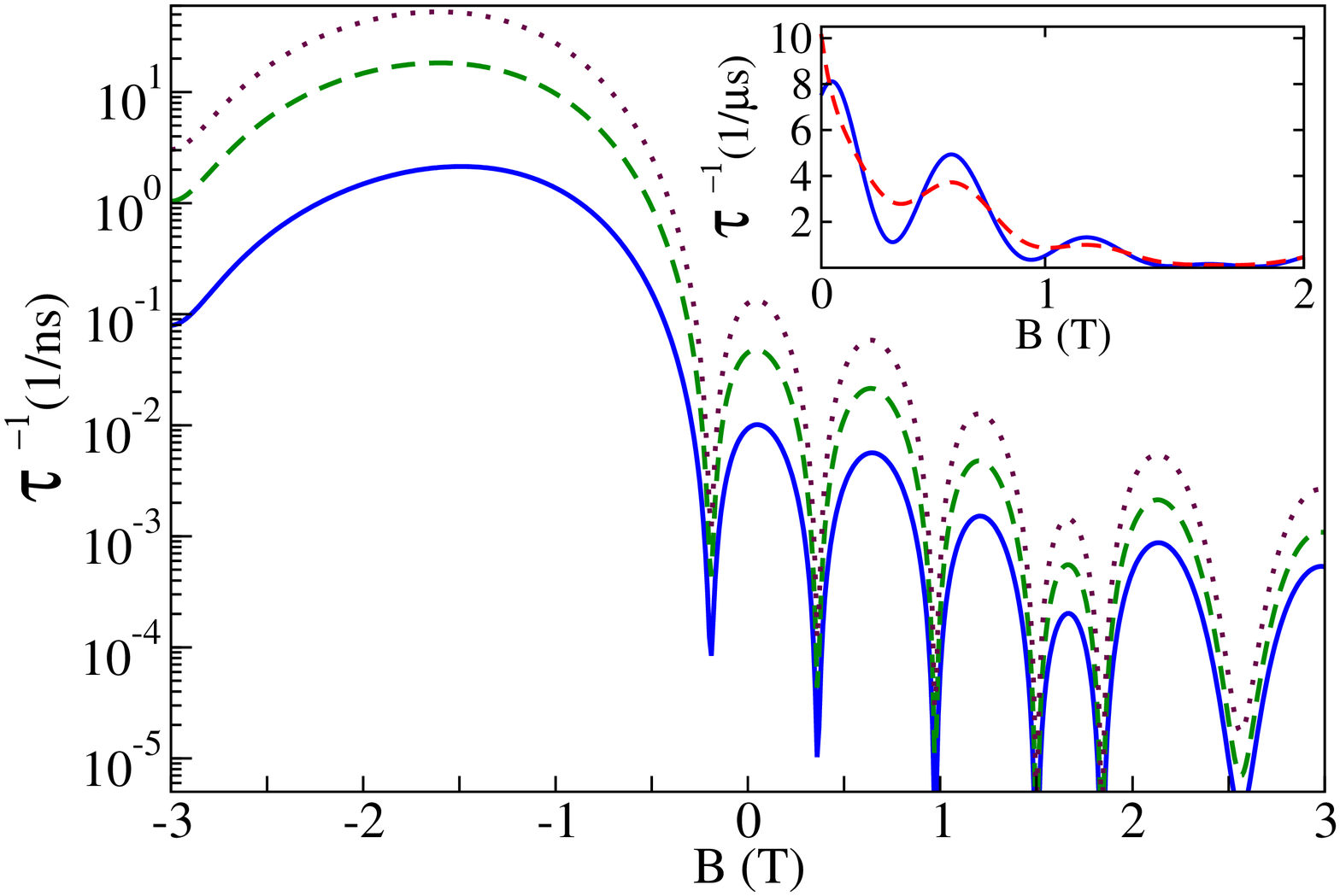}
\caption{\label{fig2} }
\end{figure}

The modulation shown in   Fig.~\ref{fig1} is obtained by  changing
the lateral confining potential, i.e.,  a structure parameter that
could be hardly tuned in an experimental setup.  We show in the
following that the strong suppression of  acoustic-phonon induced
scattering can also be driven by a magnetic field B parallel to
the growth direction.  Indeed, B induces (a) a non-uniform shift
in the energies $E_i$  and $E_f$  and, consequently, a  shift  in
the emitted phonon wavevector $q_0$; (b) a modification of the
in-plane scattering matrix $M_{fi}^{(xy)}$ originating  from the
stronger  localization of the wavefunctions.  Figure \ref{fig2}
shows that, for a CQD with a $2$~meV confining energy,  the
acoustic-phonon  scattering   rate is reduced by four orders    of
magnitude at $B = 1$~T.
%
%
In order   to  test  the   robustness  of  this  effect  in
realistic experimental conditions,  and  particularly to compare
with possible pump-and-probe experiments \cite{fujisawa02},   in
which the  electron energy could  vary  between  the cycles
leading   to a finite  energy uncertainty,  we introduced a
Gaussian smearing in the  QD confining energy.  The resulting
scattering rate (Fig.~\ref{fig2}, inset)  still shows oscillations
of about one order of magnitude when an uncertainty of 0.2~meV is
assumed.

\begin{figure}
\includegraphics[width=.98\linewidth]{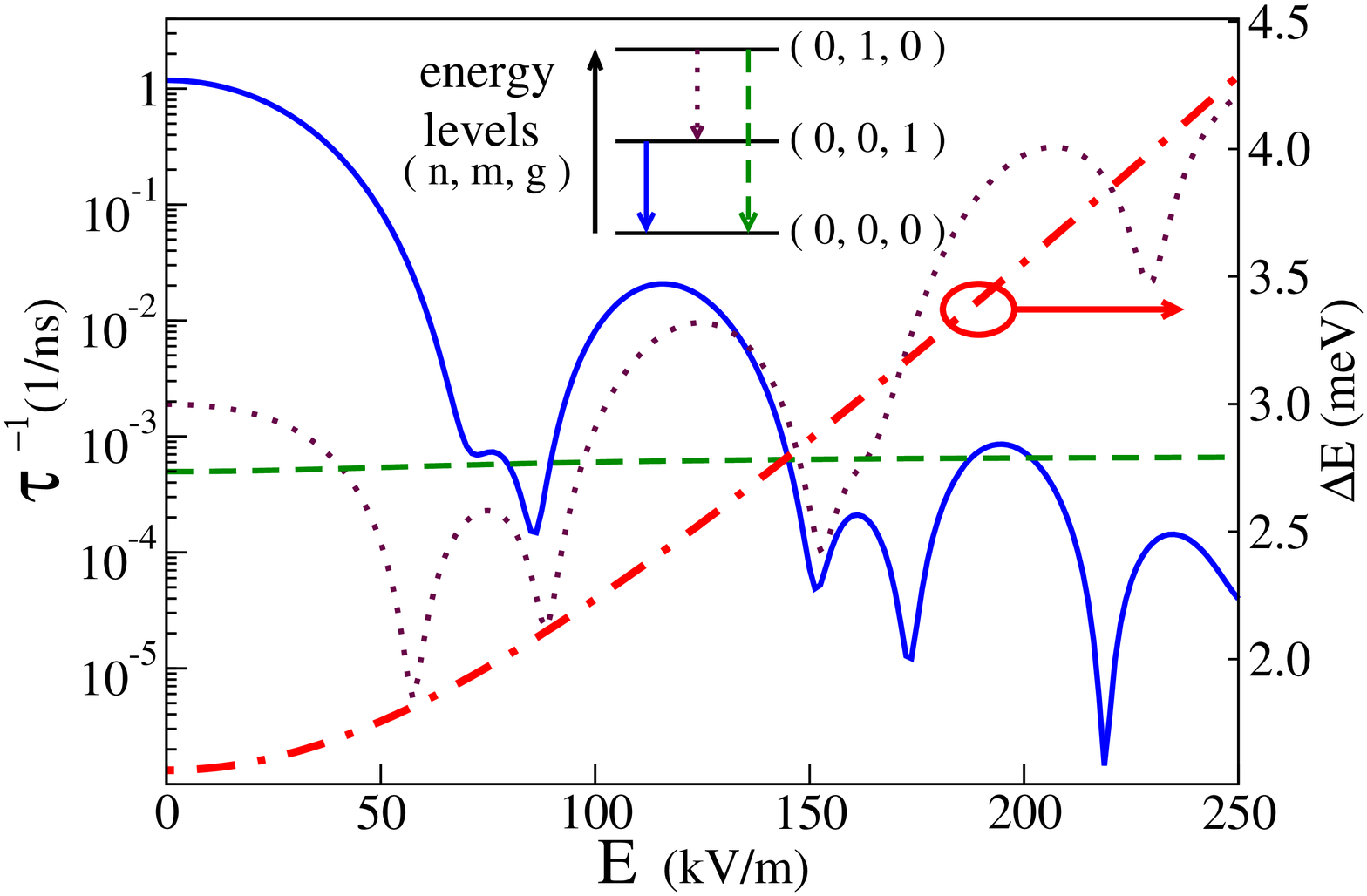}
\caption{\label{fig3} }
\end{figure}

Next, we  analyze   the   transition rate  between   two
different pseudo-spin levels, $(n,m,g)=  (0,0,1)\rightarrow
(0,0,0)$, i.e., with the $z$ component of the electron wave
function  decaying from the first excited   to    the ground  state.
We      consider a  sample     with
$\hbar\omega_0$ substantially larger than  the tunneling energy,
so that the above transition is the  lowest (see Fig.~\ref{fig3}, caption).
In this case the wavelength
of the emitted phonon is controlled by the tunneling energy, that
can be tuned by means of an  electric field applied in  the  $z$
direction.  As in the previous cases, order-of-magnitude
modulation of the  lowest transition can be achieved  at  specific
fields   (Fig.~\ref{fig3},   solid line); furthermore, scattering
rates tend to be smaller at higher fields as a result of
increasing the wavevector of  the emitted phonon, whose energy
is also shown.
Note that the effect  of the fields  on the scattering rates
generally depends  on the specific  transition, therefore
making it possible to tailor  the  scattering
rates and to make only specific states robust to   phonon-induced
decoherence. This should allow a selective population of excited states.
As an example,    in Fig.~\ref{fig3} we report the
scattering rates of the  $2^{\mbox{\scriptsize nd}}$ excited state
to the lower ones (see inset):  at  $E \sim  200$~kV/m (0,1,0)
decays preferentially   to (0,0,1)  which,  in turn,  is long
lived.
Finally, we remark that the observation of the field-dependent
oscillation of the transition rate would serve  as a signature for the
coherent delocalization of electron states in CQDs.


In summary  we have shown that  a strong suppression of LA-scattering,
the main   source of electron  decoherence   in quasi zero dimensional
systems, can be  obtained in CQDs  with energy-level separation around
3~meV.   Significantly,  the  relaxation  time  can be  controlled  by
external  magnetic or  electric  field  and increased   up to tens  of
$\mu$s; this should  be compared to  SQDs with similar level  spacing,
where scattering times   are  of the order  of   ns and not   strongly
affected by external fields.

This work  has been   partially supported by   projects
MIUR-FIRB n.RBAU01ZEML, MIUR-COFIN  n.2003020984,  INFM    Calcolo
Parallelo 2004, and MAE, Dir.Gen. Promozione Cooperazione Culturale.

\newpage


\newpage

FIGURE CAPTIONS

\vspace{1cm}

FIG. 1. Scattering rate in a CQD ($L_d=10$~nm,
$L_b=3$~nm) for the transition
$(n,m,g)=(0,-1,0)\rightarrow(0,0,0)$ as a   function   of   the
in-plane   parabolic   confinement    energy $\hbar\omega_0$. The
dotted line shows   the value of $\hbar\omega_0$ used for  the
simulations with an  applied  magnetic field. Inset: the
scattering rate for  the same sample is  compared with that for  a
CQD with   a larger  barrier  ($L_b=8$~nm, dashed    curve) and
for  a SQD ($L_d=10$~nm, dotted curve), in a linear scale.

\vspace{1cm}

FIG. 2. Scattering rate in a CQD ($L_d=10$~nm,
$L_b=3$~nm,   $\hbar\omega_0=2$~meV)  for          the      transition
$(n,m,g)=(0,-1,0)\rightarrow  (0,0,0)$ \cite{footnote4} as a  function
of the   applied  vertical magnetic field  $B$   (see  text) at  three
temperatures,  $T=0$~K  (solid),  $100$~K  (dashed), $300$~K (dotted).
The proper Bose statistics for phonon modes has been used in the
latter two simulations.
Inset:  two different   energy-uncertainty conditions are  compared at
$T=0$K:  the data are convoluted with  a Gaussian uncertainty of about
10\%  (FWHM=$0.2$~meV,  solid line) and  20\%  (FWHM=$0.4$~meV, dashed
line) on the dot confining energy.

\vspace{1cm}

FIG. 3. Scattering rate (solid line), for the
lowest pseudo-spin transition  $(n,m,g)= (0,0,1)\rightarrow
(0,0,0)$ in a CQD with $L_d=12$~nm, $L_b=4$~nm, and
$\hbar\omega_0=5$~meV as   a function of an electric  field
applied along $z$. The scattering rate for $(0,1,0)\rightarrow
(0,0,1)$  and $(0,1,0)\rightarrow  (0,0,0)$ are shown with dotted
and dashed lines  respectively. The energy of  the emitted LA
phonon is reported for reference (thick dash-dotted curve, right axis).


\newpage

\vspace{10cm}
\newpage

\vspace{10cm}
\newpage

\begin{thebibliography}{11}
\expandafter\ifx\csname natexlab\endcsname\relax\def\natexlab#1{#1}\fi
\expandafter\ifx\csname bibnamefont\endcsname\relax
  \def\bibnamefont#1{#1}\fi
\expandafter\ifx\csname bibfnamefont\endcsname\relax
  \def\bibfnamefont#1{#1}\fi
\expandafter\ifx\csname citenamefont\endcsname\relax
  \def\citenamefont#1{#1}\fi
\expandafter\ifx\csname url\endcsname\relax
  \def\url#1{\texttt{#1}}\fi
\expandafter\ifx\csname urlprefix\endcsname\relax\def\urlprefix{URL }\fi
\providecommand{\bibinfo}[2]{#2}
\providecommand{\eprint}[2][]{\url{#2}}

\bibitem{dotsrev}
\bibinfo{author}{\bibfnamefont{L.}~\bibnamefont{Jacak}},
  \bibinfo{author}{\bibfnamefont{P.}~\bibnamefont{Hawrylak}}, \bibnamefont{and}
  \bibinfo{author}{\bibfnamefont{A.}~\bibnamefont{W\'{o}js}},
  \emph{\bibinfo{title}{Quantum Dots}} (\bibinfo{publisher}{Springer, Berlin},
  \bibinfo{year}{1998});
\bibinfo{author}{\bibfnamefont{W.~G.} \bibnamefont{van~der Wiel}},
  \bibinfo{author}{\bibfnamefont{S.~D.} \bibnamefont{Franceschi}},
  \bibinfo{author}{\bibfnamefont{J.~M.} \bibnamefont{Elzerman}},
  \bibinfo{author}{\bibfnamefont{T.}~\bibnamefont{Fujisawa}},
  \bibinfo{author}{\bibfnamefont{S.}~\bibnamefont{Tarucha}}, \bibnamefont{and}
  \bibinfo{author}{\bibfnamefont{L.~P.} \bibnamefont{Kouwenhoven}},
  \bibinfo{journal}{Rev.\ Mod.\ Phys.} \textbf{\bibinfo{volume}{75}},
  \bibinfo{pages}{1} (\bibinfo{year}{2003});
\bibinfo{author}{\bibfnamefont{S.~M.} \bibnamefont{Reimann}} \bibnamefont{and}
  \bibinfo{author}{\bibfnamefont{M.}~\bibnamefont{Manninen}},
  \bibinfo{journal}{Rev.\ Mod.\ Phys.} \textbf{\bibinfo{volume}{74}},
  \bibinfo{pages}{1283} (\bibinfo{year}{2002}).






\bibitem[{\citenamefont{Bockelmann and Bastard}(1990)}]{bockelmann90}
\bibinfo{author}{\bibfnamefont{U.}~\bibnamefont{Bockelmann}} \bibnamefont{and}
  \bibinfo{author}{\bibfnamefont{G.}~\bibnamefont{Bastard}},
  \bibinfo{journal}{Phys.\ Rev.\ B} \textbf{\bibinfo{volume}{42}},
  \bibinfo{pages}{8947} (\bibinfo{year}{1990}).

\bibitem{inoshita97hameau99}
\bibinfo{author}{\bibfnamefont{T.}~\bibnamefont{Inoshita}} \bibnamefont{and}
  \bibinfo{author}{\bibfnamefont{H.}~\bibnamefont{Sakaki}},
  \bibinfo{journal}{Phys.\ Rev.\ B} \textbf{\bibinfo{volume}{56}},
  \bibinfo{pages}{R4355} (\bibinfo{year}{1997});
\bibinfo{author}{\bibfnamefont{S.}~\bibnamefont{Hameau}},
  \bibinfo{author}{\bibfnamefont{Y.}~\bibnamefont{Guldner}},
  \bibinfo{author}{\bibfnamefont{O.}~\bibnamefont{Verzelen}},
  \bibinfo{author}{\bibfnamefont{R.}~\bibnamefont{Ferreira}},
  \bibinfo{author}{\bibfnamefont{G.}~\bibnamefont{Bastard}},
  \bibinfo{author}{\bibfnamefont{J.}~\bibnamefont{Zeman}},
  \bibinfo{author}{\bibfnamefont{A.}~\bibnamefont{Lemaitre}}, \bibnamefont{and}
  \bibinfo{author}{\bibfnamefont{J.~M.} \bibnamefont{G\'{e}rard}},
  \bibinfo{journal}{Phys.\ Rev.\ Lett.} \textbf{\bibinfo{volume}{83}},
  \bibinfo{pages}{4152} (\bibinfo{year}{1999}).



\bibitem[{\citenamefont{Jacak et~al.}(2003)\citenamefont{Jacak, Machnikowski,
  Krasnyj, and Zoller}}]{jacak03}
\bibinfo{author}{\bibfnamefont{L.}~\bibnamefont{Jacak}},
  \bibinfo{author}{\bibfnamefont{P.}~\bibnamefont{Machnikowski}},
  \bibinfo{author}{\bibfnamefont{J.}~\bibnamefont{Krasnyj}}, \bibnamefont{and}
  \bibinfo{author}{\bibfnamefont{P.}~\bibnamefont{Zoller}},
  \bibinfo{journal}{Eur.\ Phys.\ J.\ D} \textbf{\bibinfo{volume}{22}},
  \bibinfo{pages}{319} (\bibinfo{year}{2003}).


\bibitem[{\citenamefont{Zanardi and Rossi}(1998)}]{zanardi98}
\bibinfo{author}{\bibfnamefont{P.}~\bibnamefont{Zanardi}} \bibnamefont{and}
  \bibinfo{author}{\bibfnamefont{F.}~\bibnamefont{Rossi}},
  \bibinfo{journal}{Phys.\ Rev.\ Lett.} \textbf{\bibinfo{volume}{81}},
  \bibinfo{pages}{4752} (\bibinfo{year}{1998}).

\bibitem[{\citenamefont{Fujisawa et~al.}(2002)\citenamefont{Fujisawa, Austing,
  Tokura, Hirayama, and Tarucha}}]{fujisawa02}
\bibinfo{author}{\bibfnamefont{T.}~\bibnamefont{Fujisawa}},
  \bibinfo{author}{\bibfnamefont{D.~G.} \bibnamefont{Austing}},
  \bibinfo{author}{\bibfnamefont{Y.}~\bibnamefont{Tokura}},
  \bibinfo{author}{\bibfnamefont{Y.}~\bibnamefont{Hirayama}}, \bibnamefont{and}
  \bibinfo{author}{\bibfnamefont{S.}~\bibnamefont{Tarucha}},
  \bibinfo{journal}{Nature} \textbf{\bibinfo{volume}{419}},
  \bibinfo{pages}{278} (\bibinfo{year}{2002}).

\bibitem{tarucha96rontani04}
\bibinfo{author}{\bibfnamefont{S.}~\bibnamefont{Tarucha}},
  \bibinfo{author}{\bibfnamefont{D.~G.} \bibnamefont{Austing}},
  \bibinfo{author}{\bibfnamefont{T.}~\bibnamefont{Honda}},
  \bibinfo{author}{\bibfnamefont{R.~J.} \bibnamefont{van~der Hage}},
  \bibnamefont{and} \bibinfo{author}{\bibfnamefont{L.~P.}
  \bibnamefont{Kouwenhoven}},
  \bibinfo{journal}{Phys.\ Rev.\ Lett}
  \textbf{\bibinfo{volume}{77}}, \bibinfo{pages}{3613} (\bibinfo{year}{1996});
\bibinfo{author}{\bibfnamefont{M.}~\bibnamefont{Rontani}},
  \bibinfo{author}{\bibfnamefont{S.}~\bibnamefont{Amaha}},
  \bibinfo{author}{\bibfnamefont{K.}~\bibnamefont{Muraki}},
  \bibinfo{author}{\bibfnamefont{F.}~\bibnamefont{Manghi}},
  \bibinfo{author}{\bibfnamefont{E.}~\bibnamefont{Molinari}},
  \bibinfo{author}{\bibfnamefont{S.}~\bibnamefont{Tarucha}}, \bibnamefont{and}
  \bibinfo{author}{\bibfnamefont{D.~G.} \bibnamefont{Austing}},
  \bibinfo{journal}{Phys.\ Rev.\ B} \textbf{\bibinfo{volume}{69}},
   \bibinfo{pages}{85327} (\bibinfo{year}{2004}).



\bibitem{footnote1}
In the numerical simulations we consider a GaAs/Al$_{0.3}$Ga$_{0.7}$As
heterostructure  and    use  the following   parameters:  $D=8.6$~eV,
$\rho=5300$~kg/m$^3$, $v=3700$~m/s, $V_h-V_l= 243$~meV.

\bibitem{footnote2}
The subscripts $i$  and $f$ indicate,  in compact notation, the  three
quantum numbers $(n_i,m_i,g_i)$ and $(n_f,m_f,g_f)$ of the initial and
final state, respectively.

\bibitem[{\citenamefont{Bockelmann}(1994)}]{bockelmann94}
\bibinfo{author}{\bibfnamefont{U.}~\bibnamefont{Bockelmann}},
  \bibinfo{journal}{Phys.\ Rev.\ B} \textbf{\bibinfo{volume}{50}},
  \bibinfo{pages}{17271} (\bibinfo{year}{1994}).



\bibitem{footnote4}
At  negative   B,   $(0,-1,0)$    is   higher  than   $(0,1,0)$;
the $(0,1,0)\rightarrow (0,0,0)$ scattering rate can be inferred
from that of $(0,-1,0)\rightarrow (0,0,0)$,   reported in
Fig.~\ref{fig2},  by changing the sign of the field B.

\end{thebibliography}
\end{document}